\documentclass[12pt,oneside]{article}
\usepackage{amsfonts,amssymb,graphicx}
\newcommand{\Av}{{\rm Av}}
\newcommand{\p}{\mathbb P}

\def\be{\begin{equation}}
\def\ee{\end{equation}}
\def\bea{\begin{eqnarray}}
\def\eea{\end{eqnarray}}
\def\c{\bar{c}}
\def\<{\langle}
\def\>{\rangle}
\def\~{\tilde}
\def\s{\sigma}

\def\b{\beta}

\def\t{\tau}

\newcommand{\qed}{\hfill \ensuremath{\Box}}

\newcommand{\Z}{\Bbb Z}

\newcommand{\av}[1]{\mbox{{\rm Av}}\left(#1\right)}

\newtheorem{remark}{Remark}

\newtheorem{theorem}{Theorem}

\newtheorem{lemma}{Lemma}

\begin{document}
\begin{center}
\vspace{1truecm}
{\bf\sc\Large spin glass identities\\
and the nishimori line}\\
\vspace{1cm}
{Pierluigi Contucci$^{\dagger}$, Cristian Giardin\`a$^{\ddagger}$, Hidetoshi Nishimori$^{\star}$}\\
\vspace{.5cm}
{\small $\dagger$ Dipartimento di Matematica} \\
{\small Universit\`a di Bologna,40127 Bologna, Italy}\\
{\small {e-mail: {\em contucci@dm.unibo.it}}}\\
\vspace{.5cm}
{\small $\ddagger$ Eindhoven University of Technology \& EURANDOM}\\
{\small P.O. Box 513 - 5600 MB Eindhoven, The Netherland}\\
{\small {e-mail: {\em c.giardina@tue.nl}}}\\
\vspace{.5cm}
{\small $\star$ Department of Physics} \\
{\small Tokyo Institute of Technology, Oh-okayama, Meguru-ku, Tokyo 152-8551, Japan}\\
{\small {e-mail: {\em nishimori@phys.titech.ac.jp}}}\\

\vskip 1truecm
\end{center}
\vskip 1truecm
\begin{abstract}\noindent
For a general spin glass model with asymmetric couplings we prove a family of
identities involving expectations of generalized overlaps and magnetizations
in the quenched state. Those identities holds pointwise in the Nishimori
line and are reached at the rate of the inverse volume while, in the 
general case, they can be proved in integral average.
\end{abstract}
\newpage\noindent

\section{Introduction and results}
Overlap identities have played and continue to play a central role in spin glass
statistical mechanics since the appearance of the Parisi solution \cite{MPV} of the Sherrington
Kirkpatrick model \cite{SK}. The replica symmetry breaking theory contains indeed, as a built-in
ansatz property, a family of identity for the overlap expectations that have been since
then classified as replica equivalence and ultrametricity factorization properties. While the first
is now largely understood for both mean field and short range finite dimensional models \cite{CGi, CGi2} the
second is still an open conjecture even within the mean field cases. The ideas to obtain the rigorous
proof of the identities trace back to the papers \cite{AC,GG} where the invariance property of stochastic
stability and the role of the energy fluctuations for the spin glass quenched measure were introduced (see also \cite{Ba}
for a different and original derivation).
The mentioned properties state a striking feature for the quenched measure on overlap expectations:
the overlap moments do obey sum rules that reduce the distributional degrees of freedom and, at
least in the mean field case, it is expected that the entire distribution be identified by the simple
overlap distribution. Those relations are expected to hold everywhere but on isolated singularities
and they can be in fact rigorously proved in $\beta$ Riemann integral average \cite{CGi2,T,B}.

We present in this paper a new family of identities in terms of generalized overlap and magnetization
that hold when the interactions are not centered which are proved in integral average. Moreover we find the
remarkable result that in the Nishimori line they hold everywhere with respect to the parameters and are reached
at the expected rate of the inverse volume.

We can illustrate the results proved in this paper considering the Edwards-Anderson
spin glass model with non symmetric Gaussian interactions of variance $\Delta^2$ and non zero average $\mu$.
Considering the quenched measure over multiple copies of the system subject to the same
disorder (for a precise definition see Section \ref{def})
the {\em link overlaps} and the {\em link magnetizations} fulfill an infinite family of identities when integrated
in the inverse temperature $\beta$ over arbitrary intervals. Those identities at the lowest moment
read, for all $\mu$ and in the thermodynamic limit:
\bea\label{cgn1}
\int_{\beta_1}^{\beta_2} d\beta  && \hspace{-.8cm}\left|\right. \beta^2  <-2q^2_{1,2}+8q_{1,2}q_{2,3}-6q_{1,2}q_{3,4}>
\nonumber \\
&+& \beta <4q_{1,2}m_3 - 4q_{1,2} m_1>
\nonumber \\
&+&  <m_1^2 - m_1m_2> \left.\right| = 0
\eea
\bea\label{cgn2}
\int_{\beta_1}^{\beta_2} d\beta  && \hspace{-.8cm}\left|\right. \beta^2 (<q_{1,2}^2> - 6<q_{1,2}><q_{2,3}> + 6 <q_{1,2}q_{3,4}> - <q_{1,2}>^2 )
\nonumber \\
&+& \beta (2<q_{1,2}m_1> - 4 <q_{1,2}m_3> + 2 <q_{1,2}><m_1>)
\nonumber \\
&+&<m_1 m_2> - <m_1>^2 \left.\right| = 0
\eea
Since we prove that the magnetization has vanishing fluctuation in $\mu$-average
in the infinite volume limit
\be
\int_{\mu_1}^{\mu_2} d\mu ( <m^2> - <m>^2) = 0
\ee
then, when integrated over arbitrary boxes in the $(\beta,\mu)$ plane, the previous identities become
\be\label{cgnr1}
\int_{\mu_1}^{\mu_2} d\mu\int_{\beta_1}^{\beta_2} d\beta|\beta^2 <-2q_{1,2}^2+8q_{1,2}q_{2,3}-6q_{1,2}q_{3,4}> | = 0
\ee
\be\label{cgnr2}
\int_{\mu_1}^{\mu_2} d\mu\int_{\beta_1}^{\beta_2} d\beta|\beta^2 ( <q_{1,2}^2> - 6<q_{1,2}><q_{2,3}> + 6 <q_{1,2}q_{3,4}> - <q_{1,2}>^2 )| = 0 \; ,
\ee
thus reducing to the standard ones \cite{Pa,G2,AC}. Notice the remarkable fact that the identities of type 
(\ref{cgn1}) have recently appeared in the theory of mean field diluted ferromagnets \cite{ABC}.

In the Nishimori line \cite{N1,N2},
\be
\beta\Delta^2=\mu
\ee
the identities (\ref{cgn1}) and (\ref{cgn2}) hold pointwise in $\beta$ and $\mu$ and the identities
(\ref{cgnr1}) and (\ref{cgnr2}) hold when integrated over $\mu$ on arbitrary intervals.

The plan of the paper is the following: in the next Section we define
the general class of models for which our result apply and we set the notations.
In Section \ref{results} we state the theorems, which are then proved
in Section \ref{proof}.

\section{Definitions}
\label{def}

We consider a disordered model of Ising configurations
$\s_n=\pm 1$, $n\in \Lambda\subset \Z^d$ for some $d$-parallelepiped
$\Lambda$ of volume $|\Lambda|$. We denote
$\Sigma_\Lambda$ the set of all $\s=\{\s_n\}_{n\in \Lambda}$, and
$|\Sigma_\Lambda|=2^{|\Lambda|}$. In the sequel the
following definitions will be used.

\begin{enumerate}

\item {\it Hamiltonian}.\\ For every $\Lambda\subset \Z^d$ let
$\{H_\Lambda(\sigma)\}_{\s\in\Sigma_N}$
be a family of
$2^{|\Lambda|}$ {\em translation invariant (in distribution)
Gaussian} random variables defined, in analogy with \cite{RU}, according to
the  general representation
\be
H_{\Lambda}(\s) \; = \; - \sum_{X\subset \Lambda} J_X\s_X
\label{hami}
\ee
where
\be
\s_X=\prod_{i\, \in X}\s_i \; ,
\ee
($\s_\emptyset=0$) and the $J$'s are independent Gaussian variables with
mean
\be
{\rm Av}(J_X) = \mu_X \; ,
\ee
and variance
\be
{\rm Av}((J_X -\mu_X)^2) = \Delta^2_X  \; .
\ee
\item {\it Average and Covariance matrix}.\\
The Hamiltonian has average
\be\label{mu}
{\cal B}_\Lambda (\s) \; := \; {\rm Av}(H_\Lambda(\s))  \;
= \; \sum_{X\subset\Lambda}\mu_X\s_X\, ,
\ee
and covariance matrix
\begin{eqnarray}
\label{cc}
{\cal C}_\Lambda (\s,\tau) \; &:= &\;
\av{[H_\Lambda(\s) - {\cal B}_\Lambda (\s)][H_\Lambda (\tau)} - {\cal B}_\Lambda (\t)])
\nonumber\\
& = & \; \sum_{X\subset\Lambda}\Delta^2_X\s_X\t_X\, .
\end{eqnarray}
By the triangular inequality
\be\label{tria}
|{\cal B}_\Lambda (\s)| \; \le \; \sum_{X\subset\Lambda}|\mu_X|
\ee
for all $\sigma$ and by the Schwarz inequality
\be\label{sw}
|{\cal C}_\Lambda (\s,\t)| \; \le \; \sqrt{{\cal C}_\Lambda
(\s,\s)}\sqrt{{\cal C}_\Lambda (\t,\t)} \; = \;
\sum_{X\subset\Lambda}\Delta^2_X
\ee
for all $\s$ and
$\t$.
\item {\it Thermodynamic Stability}.\\
The Hamiltonian (\ref{hami}) is thermodynamically stable if there exist
constants $\c$ such that
\begin{eqnarray}
\label{thst}
\sup_{\Lambda\subset\Z^d}
\frac{1}{|\Lambda|}\sum_{X\subset\Lambda}|\mu_X|
\; & \le & \; \c \; <\; \infty
\nonumber\\
\sup_{\Lambda\subset\Z^d}
\frac{1}{|\Lambda|}\sum_{X\subset\Lambda}\Delta^2_X
\; & \le & \; \c \; < \; \infty\;.
\end{eqnarray}
Thanks to the relations (\ref{tria}) and (\ref{sw}) a thermodynamically stable model fulfills the bound
\begin{eqnarray}
\label{pippo}
{\cal B}_\Lambda (\s) \; & \le & \; \c \, |\Lambda|
\nonumber\\
{\cal C}_\Lambda (\s,\t) \; & \le & \; \c \, |\Lambda|
\end{eqnarray}
and has an order $1$ normalized mean and covariance
\begin{eqnarray}
b_{\Lambda}(\s) & := & \; \frac{1}{|\Lambda|}{\cal B}_\Lambda (\s)
\nonumber\\
c_{\Lambda}(\s,\t) \; & : = & \; \frac{1}{|\Lambda|}{\cal C}_\Lambda (\s,\t)\;.
\end{eqnarray}
\item {\it Random partition function}.
\be\label{rpf}
{\cal Z}(\beta) \; := \; \sum_{\s  \in \,\Sigma_\Lambda}
e^{-\beta{H}_\Lambda(\s)}
\; .
\ee
\item {\it Random free energy}.
\be\label{rfe}
-\beta {\cal F}(\beta) \; := \; {\cal A}(\beta) \; := \; \ln {\cal Z}(\beta)
\; .
\ee
\item {\it Random internal energy}.
\be\label{rie}
{\cal U}(\beta) \; := \; \frac{\sum_{\s  \in \,\Sigma_\Lambda}
H_{\Lambda}(\s)e^{-\beta{H}_\Lambda(\s)}}{\sum_{\s  \in \,\Sigma_\Lambda}
e^{-\beta{H}_\Lambda(\s)}}
\; .
\ee
\item {\it Quenched free energy}.
\be
-\beta F(\beta) \; := \; A(\beta) \; := \; \av{ {\cal A}(\beta) }\; .
\ee
\item $R$-{\it product random Gibbs-Boltzmann state}.
\be
\Omega (-) \; := \;
\sum_{\sigma^{(1)},...,\sigma^{(R)}}(-)\,
\frac{
e^{-\beta[H_\Lambda(\s^{(1)})+\cdots
+H_\Lambda(\sigma^{(R)})]}}{[{\cal Z}(\beta)]^R}
\; .
\label{omega}
\ee
\item {\it Quenched equilibrium state}.
\be
<-> \, := \av{\Omega (-)} \; .
\ee
\item\label{obs} {\it Observables}.\\
For any smooth bounded function $G(b_{\Lambda},c_{\Lambda})$
(without loss of generality we consider $|G|\le 1$ and no assumption of
permutation invariance on $G$ is made) of the mean and covariance matrix
entries we introduce the random (with respect to $<->$) $R$-dimensional
vector of elements $\{m_k\}$ (called {\it generalized magnetization})
and the $R\times R$ matrix of elements $\{q_{k,l}\}$ (called {\it generalized
overlap}) by the formula
\be
<G(m,q)> \; := \; \av{\Omega (G(b_{\Lambda},c_{\Lambda}))} \; .
\ee
E.g.:
$G(b_{\Lambda},c_\Lambda)= b_{\Lambda}(\sigma^{1})
c_{\Lambda}(\sigma^{(1)},\sigma^{(2)})c_{\Lambda}(\sigma^{(2)}
,\sigma^{(3)})$
\be
<m_{1}q_{1,2}q_{2,3}> \; = \;
\av{\sum_{\sigma^{(1)},\sigma^{(2)},\sigma^{(3)}}
b_{\Lambda}(\sigma^{1})c_{\Lambda}(\sigma^{(1)},\sigma^{(2)})c_{\Lambda}(\sigma^{(2)},\sigma^{(3)})
\;\frac{
e^{-\beta[\sum_{i=1}^{3}H_\Lambda(\
\s^{(i)})]}}{[{\cal Z}(\beta)]^3}}
\ee
\end{enumerate}

\section{Theorems}
\label{results}

To state our results we introduce the random variable $J'$ defined by
$$
J_X = J^{'}_X + \mu_X \; ,
$$
and deform uniformly the averages $\mu_X$ with a parameter $\mu$
defined by
$$
\mu \mu^{'}_X = \mu_X \; ,
$$
in such a way that
$$
J_X = J^{'}_X + \mu \mu^{'}_X \; .
$$

Our results can be summarized in the following theorems.
\subsection{Identities in $\beta$-average}
For every observable $G$ of the kind considered in the previous
Section, we define
\begin{eqnarray}
\label{f1}
f_1(\beta,\mu)& = &
\sum_{l=1}^{R} <(m_l - m_{R+1})\,G>
+ \\
& &
 -\beta\left( <\mathop{\sum_{k,l=1}^{R}}_{k\ne l} G \,q_{\,l,\,k}
- 2R G \, \sum_{l=1}^{R} q_{\,l,\,R+1}
+R(R+1) G \, q_{\,R+1,\,R+2} > \right)
\nonumber
\end{eqnarray}
and
\begin{eqnarray}
\label{f2}
f_2(\beta,\mu)
& = &
<m_{R+1}G> - <m_1><G>
+\\
& &
-\beta \left(\sum_{k=1}^{R+1}  <G \, q_{\,k,\,R+1}> - (R+1)  <G \, q_{\,R+1,\,R+2}>
 -  <G>(<q_{1,1}>-<q_{1,2}>)
\right)
\nonumber
\end{eqnarray}

We then have the following
\begin{theorem}
\label{teo1}
The quenched equilibrium state of a thermodynamically stable Hamiltonian
fulfills, for every observable $G$ and
every temperature interval $[\b_1,\b_2]$ the following identities in the
thermodynamic limit
\begin{eqnarray}
\label{st1}
\lim_{\Lambda\nearrow\Z^d}\int_{\b_1}^{\b_2} d\b\; |f_1(\beta,\mu)| \; = \; 0
\end{eqnarray}
\begin{eqnarray}
\label{st2}
\lim_{\Lambda\nearrow\Z^d}\int_{\b_1}^{\b_2} d\b\; |f_2(\beta,\mu)| \; = \; 0
\end{eqnarray}
\end{theorem}

\vspace{.5truecm}
\begin{remark}
The two previous relations when applied to $G(m,q)=\beta q_{1,2} + m_1$ and $R=2$
yields the identities mentioned in the introduction, formulae (\ref{cgn1}) and (\ref{cgn2})
- here it is assumed that $q_{1,1}=1$, as it happens for the Edwards Anderson model.
\end{remark}

\subsection{Identities in $(\beta,\mu)$-average}
We also define
\begin{eqnarray}
\label{g1}
g_1(\beta,\mu) =
 -\beta\left( <\mathop{\sum_{k,l=1}^{R}}_{k\ne l} G \,q_{\,l,\,k}
- 2R G \, \sum_{l=1}^{R} q_{\,l,\,R+1}
+R(R+1) G \, q_{\,R+1,\,R+2} > \right)
\end{eqnarray}
and
\begin{eqnarray}
\label{g2}
g_2(\beta,\mu) &=&
-\beta \left(\sum_{k=1}^{R+1}  <G \, q_{\,k,\,R+1}> - (R+1)  <G \, q_{\,R+1,\,R+2}>
 -  <G>(<q_{1,1}>-<q_{1,2}>)
\right)\nonumber\\
& &
\end{eqnarray}
\vspace{.5truecm}
\begin{theorem}
\label{teo2}
The quenched equilibrium state of a thermodynamically stable Hamiltonian
fulfills, for every observable $G$ and
every set $[\b_1,\b_2]\times[\mu_1,\mu_2]$ the following identities in the
thermodynamic limit
\begin{eqnarray}
\label{aaa1}
\lim_{\Lambda\nearrow\Z^d}\int_{\mu_1}^{\mu_2} d\mu\;\int_{\b_1}^{\b_2} d\b\; |g_1(\beta,\mu)| \; = \; 0
\end{eqnarray}
\begin{eqnarray}
\label{aaa2}
\lim_{\Lambda\nearrow\Z^d}\int_{\mu_1}^{\mu_2} d\mu\;\int_{\b_1}^{\b_2} d\b\; |g_2(\beta,\mu)| \; = \; 0
\end{eqnarray}
\end{theorem}

\begin{remark}
Two similar families of identities have been proved in \cite{CGi2} for the centered case $\av{J_X}=0$
for all $X$. This theorem generalizes the old result and reduces to it when the observable $G$
doesn't depend on $m$. 
\end{remark}

\subsection{Identities pointwise}

\vspace{.5truecm}
\begin{theorem}
\label{teo3}
In the space of parameters $(\mu_X,\Delta_X^2)_{X\in\Lambda}$ there exists
a region called the {\it Nishimori manifold}
\be
\mu_X = \beta \Delta_X^2
\ee
where the identities of Theorem (\ref{teo1}) hold pointwise, namely
\begin{eqnarray}
\label{nnn1}
\lim_{\Lambda\nearrow\Z^d} f_1(\beta,\mu) \; = \; 0
\end{eqnarray}
\begin{eqnarray}
\label{nnn2}
\lim_{\Lambda\nearrow\Z^d} f_2(\beta,\mu) \; = \; 0
\end{eqnarray}
\end{theorem}

\section{Proofs}
\label{proof}

Theorem \ref{teo1} is proved in the lemmas of subsections \ref{ss1} and \ref{ss2}.
The proof uses only elementary methods like concentration and classical inequalities.
Along the same line Theorem \ref{teo2} is proved in subsection \ref{ss3}.
Theorem \ref{teo3} is proved in subsection \ref{ss4} making use of
an exact computation on the Nishimori manifold.

Let $h(\s)=|\Lambda|^{-1}H_\Lambda(\s)$ denote the Hamiltonian per particle. We
consider the quantity
\be
\sum_{l=1}^R \left\{<h(\s^{(l)}) \; G> - <h(\s^{(l)})><G>\right\} =
\Delta_1 G +
\Delta_2 G
\ee
where
\be
\label{delta1}
\Delta_1 G = \sum_{l=1}^R \left\{\av{\Omega[h(\s^{(l)})\,G] -
\Omega[h(\s^{(l)})]\Omega[G]} \right\}
\ee
\be
\label{delta2}
\Delta_2 G = \sum_{l=1}^R \left\{\av{\Omega[h(\s^{(l)})]\Omega[G]} -
\av{\Omega[h(\s^{(l)})]} \av{\Omega[G]}\right\}
\ee
We are going to show that both $\Delta_1 G$ and  $\Delta_2 G$
vanish (in $\beta$ average) in the thermodynamic limit.
This implies, by a simple application of integration by parts,
the relations  (\ref{st1}) and (\ref{st2}).

\subsection{Stochastic Stability Bound}
\label{ss1}
We follow the method of stochastic stability as developed in \cite{CGi}.

\begin{lemma}
\label{lnew}
For any inverse temperature interval $[\beta_1,\beta_2]$ one has
\be
\int_{\beta_1}^{\beta_2} \av{\Omega(h^2)-\Omega^2(h)} \; d\b \;\le\;
\frac{\c(2+\beta_1+\beta_2)}{|\Lambda|}
\ee
\end{lemma}
{\bf Proof.}\\
The variance of the Hamiltonian per particle $h(\sigma)$ with respect to the
Boltzmann state is nothing but (minus) the derivative of the average of $h(\sigma)$
up to a factor $1/|\Lambda|$, i.e.
\be
\label{edc}
\av{\Omega(h^2) -\Omega^2(h)} = - \frac{1}{|\Lambda|} \frac{d}{d\beta} \av{\Omega(h)}
\ee
Application of integration by parts (for its general form see formula (\ref{ibp}) below)
yields
\begin{eqnarray}
\label{u}
\av{{\Omega(h)}}
& = &
\frac{1}{|\Lambda|}\av{\sum_{X\subset \Lambda} J_X\Omega(\s_X)}
\nonumber\\
& = &
\frac{1}{|\Lambda|}\av{\sum_{X\subset\Lambda}\mu_X\Omega(\sigma_X)} + \frac{1}{|\Lambda|}\sum_{X\subset\Lambda}\b\Delta_X^2[1-\av{\Omega^2(\s_X)}]
\nonumber\\
&\le &
(1+\b)\c\;,
\end{eqnarray}
where thermodynamic stability condition, Eq. (\ref{pippo}), has been used in the last inequality.
The lemma statement follows from integration of Eq. (\ref{edc}) over an arbitrary inverse
temperature interval $[\beta_1,\beta_2]$ and the use of fundamental theorem of calculus
together with the bound (\ref{u}).
\qed
\begin{lemma}\label{l1}
For every bounded observable $G$, see definition \ref{obs} of Section \ref{def}, we have that
for every interval $[\beta_1,\beta_2]$ in the
thermodynamic limit one has
\be\label{ss}
\int_{\beta_1}^{\beta_2} |\Delta_1 G| \;d\beta \; = \; 0
\ee
\end{lemma}
{\bf Proof.}\\
From the definition of $\Delta_1 G$, Eq.(\ref{delta1}), we have
\bea
\int_{\beta_1}^{\beta_2} |\Delta_1 G| \;d\beta
& \le &
\int_{\beta_1}^{\beta_2}  \sum_{l=1}^R \;|\av{\Omega[h(\s^{(l)})\,G] -
\Omega[h(\s^{(l)})]\Omega[G]} | \; d\b
\label{abc1}
\\
& \le &
\int_{\beta_1}^{\beta_2}  \sum_{l=1}^R \;\sqrt{\av{\left\{\Omega[h(\s^{(l)})\,G] -
\Omega[h(\s^{(l)})]\Omega[G]\right\}^2} }\; d\b
\label{abc2}
\\
& \le &
R \int_{\beta_1}^{\beta_2}  \;\sqrt{\av{\Omega[h^2(\s)] -
\Omega^2[h(\s)]} }\; d\b
\label{abc3}
\\
& \le &
R \sqrt{\beta_2-\beta_1} \sqrt{\int_{\beta_1}^{\beta_2}  \;\av{\Omega[h^2(\s)] -
\Omega^2[h(\s)]} \; d\b}
\label{abc4}
\\
& \le &
R \sqrt{\beta_2-\beta_1} \sqrt{\frac{\c(2+\beta_1+\beta_2)}{|\Lambda|}}
\label{abc5}
\eea
where (\ref{abc1}) follows from triangular inequality, (\ref{abc2}) is obtained by
applying Jensen inequality on the measure $\av{-}$, (\ref{abc3}) comes from application
of Schwarz inequality to the measure $\Omega(-)$ and boundedness of $G$, (\ref{abc4})
is again Jensen inequality on the measure $\frac{1}{\beta_2-\beta_1}\int_{\beta_1}^{\beta_2}(-)d\b$
and finally (\ref{abc5}) comes from lemma \ref{lnew}.
%
\qed
\begin{remark}
The previous lemma is related to a general property of
disordered systems which is known as stochastic stability
(see \cite{AC,CGi}). It says that the equilibrium state in a spin glass
model is invariant under a suitable class of perturbation in all
temperature intervals of continuity. The result presented in (\cite{CGi})
for the case of zero average couplings holds with the absolute value outside the integral 
in beta with a vanishing rate of the inverse volume. Here instead we got the 
stronger result for the absolute value inside the integral but with a weaker
vanishing rate of the square root inverse volume. 
\end{remark}
\vspace{0.3cm}
\noindent
\begin{lemma}\label{l2} The following expression holds:
\begin{eqnarray}
\label{expr-delta1}
\Delta_1 G \; & = & \;
\sum_{l=1}^{R} <(m_l - m_{R+1})\,G>
\nonumber \\
& - & \b < G\,\left[ \mathop{\sum_{k,l=1}^{R}}_{k\ne l}  \,q_{\,l,\,k}
- 2R  \, \sum_{l=1}^{R} q_{\,l,\,R+1}
+R(R+1) \, q_{\,R+1,\,R+2} \right] > \; .
\end{eqnarray}
\end{lemma}
{\bf Proof.}\\
For each replica $l$ $(1\le l\le R)$, we evaluate separately
the two terms in the sum of the right side of Eq. (\ref{delta1}) by using
the integration by parts (generalized Wick formula) for
correlated Gaussian random variables, $x_1,x_2,\ldots,x_n$
with means $\av{x_i}$ and covariances $\av{(x_i-\av{x_i})(x_j-\av{x_j})}$,
namely
\bea\label{ibp}
\av { x_i\, \psi(x_1,\ldots,x_n) }
& = &
\av{x_i}\av{\psi(x_1,\ldots,x_n)}  \\
& + &
\sum_{j=1}^n \av{(x_i-\av{x_i})(x_j-\av{x_j})} \,
\av {\frac{\partial \psi(x_1,...,x_n)}{\partial x_j} } \;.\nonumber
\eea
It is convenient to denote by $p\,(R)$ the Gibbs-Boltzmann weight
of $R$ copies of the system
\be
p_R\,(\s^1,\ldots\s^R) \,= \,
\frac{
e^{-\beta\,[\,\sum_{k=1}^R H_\Lambda(\s^{(k)})\,]}}
{[{\cal Z}(\beta)]^R} \;,
\ee
so that we have
\be
\label{derivataBolt}
- \frac{1}{\beta}\frac{dp_R\,(\s^1,\ldots,\s^R)}{dH(\tau)} \;=\;
p_R\,(\s^1,\ldots,\s^R)
\left(\sum_{k=1}^R \delta_{\s^{(k)},\,\tau}\right)
- R \;p_{R+1}\,(\s^1,\ldots,\s^R,\tau) \;.
\ee
We obtain
\bea
\av{\Omega(h(\s^{(l)})\,G)} & = &
\frac{1}{|\Lambda|}\,
\av{\;
\sum_{\sigma^{(1)},...,\sigma^{(r)}}\;
G\;H_{\Lambda}(\s^{(l)}) \;
p_R\,(\s^1,\ldots,\s^R)} \nonumber\\
& = &
\av{ \;
\sum_{\sigma^{(1)},...,\sigma^{(r)}} \;G\;
{b}_{\Lambda}(\s^{(l)})\;p_R\,(\s^1,\ldots,\s^R)}
\qquad\qquad
\nonumber\\
& &
\label{line1}
 +\, \av{ \;
\sum_{\sigma^{(1)},...,\sigma^{(r)}}\;\sum_{\tau}\;G\;
{c}_{\Lambda}(\s^{(l)},\tau)\;
\frac{dp_R\,(\s^1,\ldots,\s^R)}{dH(\tau)}}
\qquad\qquad
\\
& = &
\label{line2}
<m_l\,G>- \beta \, \left[
\sum_{k=1}^{R} <G \,q_{\,l,\,k}> -
R  <G \, q_{\, l ,\, R+1}>
\right]
\eea
where in (\ref{line1}) we made use of the integration
by parts formula and (\ref{line2}) is obtained
by (\ref{derivataBolt}).
Analogously, the other term reads
\bea
\;
\av{\Omega(h(\s^{(l)}))\,\Omega (G)} & = &
\frac{1}{|\Lambda|}\,
\av{\;
\sum_{\sigma^{(l)}}\sum_{\tau^{(1)},...,\tau^{(R)}}\;
G\;H_{\Lambda}(\s^{(l)}) \;
p_{R+1}\,(\s^l,\tau_1,\ldots,\tau_R)}
\nonumber \\
& = &
\av{\;
\sum_{\sigma^{(l)}}\sum_{\tau^{(1)},...,\tau^{(R)}}\;G\;
{b}_{\Lambda}(\s^{(l)})\;
p_{R+1}\,(\s^l,\tau_1,\ldots,\tau_R)}\quad
\qquad
\nonumber\\
& &
\label{line3}
+\, \av{\;
\sum_{\sigma^{(l)}}\sum_{\tau^{(1)},...,\tau^{(R)}}\;\sum_{\gamma}\;G\;
{c}_{\Lambda}(\s^{(l)},\gamma)\;
\frac{dp_{R+1}\,(\s^l,\tau_1,\ldots,\tau_R)}{dH(\gamma)}}\quad
\qquad
\nonumber\\
& = &
\label{line4}
<m_{R+1} \,G>\,- \beta\,\left [
\sum_{k=1}^{R+1} <G \, q_{\,k,\,R+1}>
- (R+1) <G \, q_{\,R+1,\,R+2}>
\right]\nonumber\\
\eea
Inserting the (\ref{line2}) and (\ref{line4}) in Eq. (\ref{delta1})
and summing over $l$ we obtain the expression (\ref{expr-delta1}).
\qed
\subsection{Selfaveraging Bound}\label{ss2}

The selfaveraging of the free energy is a well established
property of spin glass models. The vanishing of the fluctuations
with respect to the disorder of the free energy can be obtained
either by martingales arguments \cite{PS,CGi2} or by concentration
of measure \cite{T,GT2}.
Here we follow the second approach. Our formulation applies to both
mean field and finite dimensional models and, for instance, includes
the non summable interactions in finite dimensions \cite{KS} and the
$p$-spin mean field model as well as the REM  and GREM  models.
\begin{lemma}
\label{martin}
The disorder fluctuation of the free energy satisfies the following inequality: for all $x > 0$
\be
\label{concentra}
\p\,\left(|{\cal A}-\av{{\cal A}}| \ge x\right) \;\le\; 2 \exp{\left(-\frac{x^2}{2 \c |\Lambda|}\right)}
\ee
The free energy is then a self averaging quantity, i.e.
\be\label{sav}
V({\cal A}) \; = \;  \av{{\cal A}^2}-\av{{\cal A}}^2 \;\le\; 4\, \c\, \beta^2\, {|\Lambda|}
\ee
\end{lemma}
{\bf Proof.}
Consider an $s > 0$. By Markov inequality, one has
\bea
\label{markov}
\p\,\left\{{\cal A}-\av{{\cal A}} \ge x\right\}
& = &
\p\,\left\{\exp [s({\cal A}-\av{{\cal A}})] \ge \exp (sx)\right\}
\nonumber\\
&\le & \av{\exp[s({\cal A}-\av{{\cal A}})]} \; \exp(-sx)
\eea
To bound the generating function
\be
\av{\exp[s({\cal A}-\av{{\cal A}})]}
\ee
one introduces, for a parameter $t\in [0,1]$, the following interpolating function:
\be
\phi(t) = \ln Av_1\{\exp(s \; Av_2\{\ln {Z}(t) \})\}\;,
\ee
where $Av_1\{-\}$ and  $Av_2\{-\}$ denote expectation with
respect to two independent copies $X_1(\s)$ and $X_2(\s)$
of the random variable $X(\s)$, which is a {\em centered} Gaussian process
with the same covariance as the Hamiltonian $H_{\Lambda}(\s)$, and the partition function ${Z}(t)$ is
\be
{Z}(t) \; = \; \sum_{\s  \in \,\Sigma_\Lambda}
e^{-\b\sqrt{t} X_1(\s) - \beta\sqrt{1-t} X_2(\s) - \b {\cal B}_{\Lambda}(\s)}
\; .
\ee
Indeed, since $H_{\Lambda}(\s) = X(\s) + {\cal B}_{\Lambda}(\s)$, it is immediate to verify (see definition (\ref{rfe})) that
\be
\phi(0) = s \;\av{\cal A}  \;,
\ee
and
\be
\phi(1) = \ln \av{e^{s \;{\cal A}}} \;.
\ee
This implies that
\be
\av{\exp[s({\cal A}-\av{{\cal A}})]} = e^{\phi(1)-\phi(0)} = e^{\int_{0}^1 \phi'(t) dt}
\ee
On the other hand, the derivative with respect to $t$ can be easily bounded. Defining
\be
K(t) = \exp(s \; Av_2\{\ln {Z}(t) \})
\ee
and
\be
p^t(\s) = \frac{e^{-\b\sqrt{t} X_1(\s) - \beta\sqrt{1-t} X_2(\s) - \b {\cal B}_{\Lambda}(\s)}}{Z(t)}
\ee
one has
\be
\phi'(t) =
\frac{Av_1\Big\{K(t)
\; s \; Av_2\left\{ \sum_{\s} \; p^t(\s)\left[\frac{1}{2\sqrt{t}} X_1(\s) - \frac{1}{2\sqrt{1-t}}X_2(\s)\right] \right\}   \Big\} }
{Av_1\{K(t)\}}
\ee
Applying the integration by parts formula (\ref{ibp}), a simple computation gives
\bea
\sum_{\s} Av_1\left\{K(t)
\; Av_2\left\{ p^t(\s)\frac{1}{\sqrt{t}}\;X_1(\s) \right \}   \right\}
& = &
s \sum_{\s,\tau}Av_1\left\{K(t) \; {\cal C}_{\Lambda}(\s,\tau)\; Av_2\left\{ p^t(\tau) \right \}
\; Av_2\left\{ p^t(\s) \right \}   \right\}
\nonumber\\
&+&
Av_1\left\{K(t)
 \; Av_2\left\{\sum_{\s}{\cal C}_{\Lambda}(\s,\s) p^t(\s)  \right\}   \right\}
\nonumber\\
& - &
Av_1\left\{K(t)
 \; Av_2\left\{\sum_{\s,\tau}{\cal C}_{\Lambda}(\s,\tau) p^t(\s)p^t(\tau) )   \right\}   \right\}
 \nonumber
\eea
and
\bea
Av_1\left\{K(t)
 \; Av_2\left\{\sum_{\s} p^t(\s)\frac{1}{\sqrt{1-t}} \;X_2(\s)  \right\}   \right\}
&= &
Av_1\left\{K(t)
 \; Av_2\left\{\sum_{\s}{\cal C}(\s,\s)_{\Lambda} p^t(\s)  \right\}   \right\}
\nonumber\\
& - &
Av_1\left\{K(t)
 \; Av_2\left\{\sum_{\s,\tau}{\cal C}(\s,\tau)_{\Lambda} p^t(\s)p^t(\tau) )   \right\}   \right\}
\nonumber
\eea
Taking the difference of the previous two expressions one finds
\be
\phi'(t) =\frac{s^2}{2}
\frac{\sum_{\s,\tau}Av_1\left\{K(t) \; {\cal C}_{\Lambda}(\s,\tau)\; Av_2\left\{ p^t(\tau) \right \}
\; Av_2\left\{ p^t(\s) \right \}   \right\}}
{Av_1\{K(t)\}}
\ee
Using the thermodynamic stability condition (\ref{pippo}), this yields
\be
|\phi'(t)| \le \frac{s^2}{2}|\Lambda| \c
\ee
from which it follows
\be
\av{\exp[s({\cal A}-\av{{\cal A}})]} \le \exp\left({\frac{s^2}{2}|\Lambda| \c}\right)
\ee
Inserting this bound into the inequality (\ref{markov}) and optimizing over $s$
one finally obtains
\be
\p\,\left({\cal A}-\av{{\cal A}} \ge x\right) \;\le\;\exp{\left(-\frac{x^2}{2 \c |\Lambda|}\right)}
\ee
The proof of inequality (\ref{concentra}) is completed by observing
that one can repeat a similar computation for  $\p\,\left({\cal A}-\av{{\cal A}} \le -x\right)$.
The result for the variance (\ref{sav}) is then immediately proved
using the identity
\be
\av{({\cal A} - \av{{\cal A}})^2} = 2 \int_{0}^{\infty} x\; \p(|{\cal A}-\av{{\cal A}}| \ge x) \;dx
\ee

\qed
\begin{lemma}\label{sa}
The internal energy is self averaging almost everywhere in $\b$, i.e. defining
$u={\cal U}/|\Lambda|$ and
$V(u)=\av{u^2}-\av{u}^2$ it holds in the thermodynamic limit
\be\label{vtoz}
\int_{\beta_1}^{\beta_2} V(u) \, d\b \; \to \; 0
\ee
\end{lemma}
{\bf Proof.}\\
The result is obtained in two steps which use general theorems of measure
theory. First from lemma \ref{martin} we obtain the
convergence to zero almost everywhere (in $\b$) of the variance of the
internal energy, then thanks to a bound on the
variance of the internal energy we apply the Lebesgue dominated convergence
theorem which gives the lemma statement.
The sequence of convex functions ${\cal A}(\beta)/|\Lambda|$ converges a.e.
(in $J$) to the limiting value $a(\beta)$ of its
average and the convergence is self averaging in the sense of lemma
\ref{martin}. By general convexity arguments \cite{RU}
it follows that the sequence of the derivatives ${\cal A}'(\beta)/|\Lambda|$
converges to $u(\beta)=a'(\beta)$ almost everywhere
in $\beta$ and also that the convergence is self averaging. In fact the
vanishing of the variance of a sequence
of convex functions is inherited, in all points in which the derivative
exists (which is almost everywhere for a convex function),
to the sequence of its derivatives (see \cite{S,OTW}). From lemma
\ref{martin} we have then
\be
V(u) \; \to \; 0 \quad \beta \;-\;  a.e.
\ee
In order to obtain the convergence in $\beta$-average we use the Lebesgue
dominated convergence theorem. In fact
we prove that the sequence of variances of $u$ is uniformly bounded (in
every interval $[\b_1,\b_2]$) by an integrable
function of $\beta$.
A lengthy but simple computation which uses again integration by parts gives
\bea\label{u2}
\av{{\cal U}^2} \; &=& \; \av{\sum_{X,Y\subset \Lambda}
J_XJ_{Y}\Omega(\s_X)\Omega(\s_Y)} \\\nonumber
\; &=& \;
\sum_{X,Y\subset\Lambda} \mu_X\mu_Y \Omega(\s_X)\Omega(\s_Y) + \sum_{X\subset\Lambda} \Delta_X^2 \Omega^2(\s_X)
\nonumber \\
& + & 2\b\sum_{X,Y\subset\Lambda} \mu_X \Delta_Y^2 \left[\Omega(\s_X\s_Y)\Omega(\s_Y)+\Omega(\s_X)-2\Omega(\s_X)\Omega^2(\s_Y)\right]
\nonumber\\
&+&
\beta^2\sum_{X,Y\subset\Lambda}\Delta_X^2\Delta_Y^2 {\rm
Av}\left[1-\Omega^2(\s_X)-\Omega^2(\s_Y)+6\Omega^2(\s_X)\Omega^2(\s_Y)+\right.\nonumber\\
& &\left. -6 \Omega(\s_X)\Omega(\s_Y)\Omega(\s_X\s_Y)+\Omega^2(\s_X\s_Y)\right]
\eea
from which
\be
V(u)\; \le \; |\Lambda|^{-2} \av{{\cal U}^2} \; \le \;  \c^2 (2 + 4\b + 14 \b^2) .
\ee
From this follows (\ref{vtoz}). \qed
\begin{lemma}\label{d2}
For every bounded observable $G$, see definition \ref{obs} of Section \ref{def}, we have
that for every interval $[\beta_1,\beta_2]$ in the thermodynamic
limit
\be
\label{stat}
\int_{\beta_1}^{\beta_2}|\Delta_2 G| \, d\beta \; = \; 0
\ee
\end{lemma}
{\bf Proof}.\\
From the definition of $\Delta_2 G$, Eq.(\ref{delta2}), we have
\bea
\int_{\beta_1}^{\beta_2} |\Delta_2 G| \;d\beta
& \le &
\int_{\beta_1}^{\beta_2}  \sum_{l=1}^R \;| \av{\Omega[h(\s^{(l)})]\Omega[G]} -
\av{\Omega[h(\s^{(l)})]} \av{\Omega[G]} | \; d\b
\label{bbc1}
\\
& \le &
\int_{\beta_1}^{\beta_2}  \sum_{l=1}^R \; \sqrt{\av{\Omega[h^2(\s^{(l)})]} -
\left(\av{\Omega[h(\s^{(l)})]}\right)^2} \; d\b
\label{bbc2}
\\
& \le &
R \int_{\beta_1}^{\beta_2}  \sqrt{V(u)} \; d\b
\label{bbc3}
\\
& \le &
R \sqrt{\beta_2 -\beta_1} \sqrt{\int_{\beta_1}^{\beta_2}  V(u) \; d\b }
\label{bbc4}
\eea
where (\ref{bbc1}) follows from triangular inequality, (\ref{bbc2}) is obtained by
applying Schwarz inequality to the measure $\av{-}$ and boundedness of $G$, (\ref{bbc4})
is Jensen inequality on the measure $\frac{1}{\beta_2-\beta_1}\int_{\beta_1}^{\beta_2}(-)d\b$.
The statement (\ref{stat}) follows then using the result of the previous lemma.
\qed
\begin{lemma} The following expression holds:
\bea\label{expr-delta2}
\Delta_2 G
& = &
\sum_{l=1}^R\left(<m_{R+1}G> - <m_l><G>\right)
+\\
& &
- \beta\, R \left [
\sum_{k=1}^{R+1}  <G \, q_{\,k,\,R+1}> - (R+1)  <G \, q_{\,R+1,\,R+2}>
-  <G>(<q_{1,1}> - <q_{1,2}>)
\right] \; . \nonumber
\eea
\end{lemma}
{\bf Proof.}
In order to obtain $\Delta_2G$ we are left with the explicit evaluation
of
the other term in (\ref{delta2}) which simply gives
\bea\nonumber
\av{\Omega(h(\s^{(l)}))}\,\av{\Omega (G)}
& = &
\frac{1}{|\Lambda|}\,
\av{\;
\sum_{\sigma^{(l)}}\;
H_{\Lambda}(\s^{(l)}) \;
p_1\,(\s)} \,<G>\\ \nonumber
& = &
\av{\;
\sum_{\sigma^{(l)}}\;
b_{\Lambda}(\s^{(l)}) \;
p_{1}\,(\s)} \,<G>\\ \nonumber
& &
+ \av{\;
\sum_{\sigma^{(l)}}\;\sum_{\gamma}\;
{c}_{\Lambda}(\s^{(l)},\gamma)\;
\frac{dp_{1}\,(\s)}{dH_{\Lambda}(\gamma)}}
\,<G>
\quad
\qquad
\\
& = &
\label{temp2}
<m_l>\,<G> - \beta\, <G>[<q_{1,1}>-<q_{1,2}>]
\eea

\noindent
Inserting the (\ref{line4}) and (\ref{temp2}) in Eq. (\ref{delta2})
and summing over $l$ we obtain the (\ref{expr-delta2}).
\qed

\subsection{Vanishing fluctuations of the generalized magnetization}
\label{ss3}

\begin{lemma} For every interval $[\mu_1,\mu_2]$, in the thermodynamic
limit
\be
\int_{\mu_1}^{\mu_2} d\mu ( <m^2> - <m>^2) = 0
\ee
\end{lemma}
{\bf Proof.}
The proof that the generalized magnetization has vanishing fluctuation
follows the strategy that has been pursued so far to control
fluctuation of the internal energy. We have
\bea
\label{uno}
<m^2> - <m>^2
& = &
\av{\Omega(b_{\Lambda}(\sigma)^2) - (\Omega(b_{\Lambda}(\sigma))^2} \\
\label{due}
& + &
\av{(\Omega(b_{\Lambda}(\sigma))^2} - (\av{\Omega(b_{\Lambda}(\sigma)})^2
\eea
and we observe that generalized magnetization is related to
the pressure by
\be
\label{jjj}
\frac{\partial }{\partial \mu}\left(\frac{{\cal A}}{|\Lambda|}\right) =
\frac{\beta}{\mu}\frac{1}{|\Lambda|}\Omega(\sum_{X\in\Lambda}\mu_X\sigma_X)
= \frac{\beta}{\mu}\;\Omega(b_{\Lambda}(\sigma))
\ee
The fluctuations w.r.t. the Gibbs state (r.h.s of Eq.(\ref{uno})) are easily controlled
by a stochastic stability argument:
\be
\label{u1}
\int_{\mu_1}^{\mu_2} d\mu \av{\Omega(b_{\Lambda}(\sigma)^2) - (\Omega(b_{\Lambda}(\sigma))^2}
=
\frac{1}{|\Lambda|}
\int_{\mu_1}^{\mu_2} d\mu \;\frac{\mu^2}{\beta^2}\;\av{\frac{\partial}{\partial\mu} \frac{\beta}{\mu}\Omega(b_{\Lambda}(\sigma))} \; ,
\ee
where the right hand side can be bounded, integrating by parts in $\mu$ by $\beta^{-1}3(\mu_2-\mu_1)$.

The fluctuations w.r.t. the disorder (Eq.(\ref{due})) are bounded by the same argument of lemma \ref{sa}.
Indeed from self-averaging of the pressure per particle
$$
V\left(\frac{ {\cal A}}{|\Lambda|}\right) \le \frac{c}{|\Lambda|} \to 0 \;,
$$
and convexity of finite volume pressure
$$
\frac{\partial^2 }{\partial \mu^2} \frac{{\cal A}}{|\Lambda|}
= {|\Lambda|} [\Omega(b_{\Lambda}(\sigma)^2) - (\Omega(b_{\Lambda}(\sigma))^2 ]\ge 0
$$
one deduces that also the sequence of derivatives (\ref{jjj}) is self-averaging
in $\mu$-average. Hence
\be
\label{u2}
\int_{\mu_1}^{\mu_2} d\mu \left[ \av{(\Omega(b_{\Lambda}(\sigma))^2} - (\av{\Omega(b_{\Lambda}(\sigma)})^2\right] =
\int_{\mu_1}^{\mu_2} d\mu V\left(\frac{\partial }{\partial \mu}\left(\frac{{\cal A}}{|\Lambda|}\right)\right) \to 0
\ee
Combining together (\ref{u1}) and  (\ref{u2}) complete the proof of the lemma.
\qed

\subsection{Nishimori manifold}
\label{ss4}
Our strategy here is to prove the vanishing of fluctuations of the Hamiltonian per particle
$h(\sigma)$.
This implies, following the same reasoning of the previous sections,
the vanishing of correlations with a generic bounded function $G$ using
the Schwarz inequality:
\be
|<hG> -<h><G>| \le \sqrt{<h^2>-<h>^2} \sqrt{<G^2>-<G>^2}\;.
\ee
\begin{lemma}
On the Nishimori manifold
\be
\mu_X = \beta \Delta_X^2
\ee
the random internal energy per particle is selfaveraging.
More precisely the following result holds:
\be
\label{var}
<h^2>-<h>^2 \; \le \; \frac{\bar{c}}{|\Lambda|}
\ee
\end{lemma}
{\bf Proof.}
This is proved by an explicit computation of both terms in formula (\ref{var}).
Let us first consider
$$
\Av \left(\Omega (J_X \sigma_X)\right)\;.
$$
We first write the definition explicitly:
\begin{equation}
 \Av \left(\Omega (J_X \sigma_X)\right)
  =\int_{-\infty}^{\infty} \prod_Y \left( dJ_Y \frac{1}{\sqrt{2\pi}\Delta_Y}
    \exp \left(-\frac{(J_Y-\mu_Y)^2}{2\Delta_Y^2} \right) \right)
    \cdot
     \frac{\sum_\sigma J_X \sigma_X e^{\beta \sum_Z J_Z \sigma_Z}}
     {\sum_\sigma e^{\beta \sum_Z J_Z \sigma_Z}}
 \label{Avdef}
\end{equation}
We apply the gauge transformation
$$
J_X\to J_X \tau_X,\quad \sigma_i \to\sigma_i \tau_i,
$$
for all $i\in\Lambda$ and $X\subset\Lambda$,
where $\tau_i$ is a `gauge' variable fixed to 1 or -1 at each $i\in X$
and $\tau_X = \prod_{i\in X}\tau_i$.
This change of variables leaves the integral and sums in the above equation invariant.
Then, only the $J_Y$ in the exponent for the Gaussian weight changes:
\begin{eqnarray}
&&\Av \left(\Omega (J_X \sigma_X)\right)\\
  &&=\int_{-\infty}^{\infty} \prod_Y \left( dJ_Y \frac{1}{\sqrt{2\pi}\Delta_Y}
    \exp \left(-\frac{(J_Y\tau_Y-\mu_Y)^2}{2\Delta_Y^2} \right) \right)
    \cdot
     \frac{\sum_\sigma J_X \sigma_X e^{\beta \sum_Z J_Z \sigma_Z}}
     {\sum_\sigma e^{\beta \sum_Z J_Z \sigma_Z}}\nonumber \\
    &&=
 \int_{-\infty}^{\infty} \prod_Y \left( dJ_Y \frac{1}{\sqrt{2\pi}\Delta_Y}
    \exp \left(-\frac{J_Y^2+\mu_Y^2}{2\Delta_Y^2} \right) \right)
    \exp \left( \sum_Y \frac{J_Y\mu_Y\tau_Y}{\Delta_Y^2} \right)
    \cdot
     \frac{\sum_\sigma J_X \sigma_X e^{\beta \sum_Z J_Z \sigma_Z}}
     {\sum_\sigma e^{\beta \sum_Z J_Z \sigma_Z}}\nonumber
\end{eqnarray}
Since this expression holds for any assignment of $\pm 1$ to $\tau_i$,
we may sum it up over all possible $\{\tau_i\}_i$ and divide the
result by $2^{|\Lambda|}$,
\begin{eqnarray}
 &&\Av \left(\Omega (J_X \sigma_X)\right)\\
 &&=\frac{1}{2^{|\Lambda|}}\int_{-\infty}^{\infty} \prod_Y
 \left( dJ_Y \frac{1}{\sqrt{2\pi}\Delta_Y}
    \exp \left(-\frac{J_Y^2+\mu_Y^2}{2\Delta_Y^2} \right) \right)
    \sum_{\tau} e^{\sum_Y J_Y\mu_Y\tau_Y/\Delta_Y^2}\nonumber\\
  &&  \cdot
     \frac{\sum_\sigma J_X \sigma_X e^{\beta \sum_Z J_Z \sigma_Z}}
     {\sum_\sigma e^{\beta \sum_Z J_Z \sigma_Z}}\nonumber
\end{eqnarray}
The sum over $\tau$ and the sum over $\sigma$ in the denominator cancel each
other for NL ($\beta =\mu_Y/\Delta_Y^2$), and we have a simplified expression
\begin{eqnarray}
 &&\Av \left(\Omega (J_X \sigma_X)\right)\\
 &&=\frac{1}{2^{|\Lambda|}}\int_{-\infty}^{\infty} \prod_Y
 \left( dJ_Y \frac{1}{\sqrt{2\pi}\Delta_Y}
    \exp \left(-\frac{J_Y^2+\mu_Y^2}{2\Delta_Y^2} \right) \right)
    \sum_\sigma J_X \sigma_X e^{\beta \sum_Z J_Z \sigma_Z}\nonumber\\
   &&=\frac{1}{2^{|\Lambda|}}\sum_\sigma \int_{-\infty}^{\infty} \prod_Y
 \left( dJ_Y \frac{1}{\sqrt{2\pi}\Delta_Y}
    \exp \left(-\frac{J_Y^2+\mu_Y^2}{2\Delta_Y^2} \right) \right)
    J_X \sigma_X e^{\ \sum_Z J_Z \sigma_Z\mu_Z/\Delta_Z^2}\nonumber
\end{eqnarray}
For given $\{\sigma_i\}_i$, let us change the integral variable as
$J_Y\to J_Y \sigma_Y$. Then $\sigma$ disappears completely and the
integral is just for the average of $J_X$
\begin{eqnarray}
 &&\Av \left(\Omega (J_X \sigma_X)\right)\nonumber\\
 &&=\frac{1}{2^{|\Lambda|}}\sum_\sigma \int_{-\infty}^{\infty} \prod_Y
 \left( dJ_Y \frac{1}{\sqrt{2\pi}\Delta_Y}
    \exp \left(-\frac{J_Y^2+\mu_Y^2}{2\Delta_Y^2} \right) \right)
    J_X \, e^{\ \sum_Z J_Z\mu_Z/\Delta_Z^2 }\nonumber\\
  &&=\frac{1}{2^{|\Lambda|}}\cdot 2^{|\Lambda|} \cdot 1 \cdot
  \int_{-\infty}^{\infty} dJ_X \frac{1}{\sqrt{2\pi}\Delta_X}
   J_X \exp \left(-\frac{(J_X-\mu_X)^2}{2\Delta_X^2}\right)\nonumber\\
   &&=\Av (J_X)
 \end{eqnarray}

From the previous computation we obtain the final result for the
quenched internal energy on the NL:
\begin{equation}
\label{media}
 \Av \left(\Omega (H_\Lambda (\sigma))\right)
 = \Av \left(\Omega \left(\sum_X J_X \sigma_X \right)\right)
 =\Av \left(\sum_X J_X \right) =\sum_X \mu_X
\end{equation}
The other term in the variance (\ref{var}) is
evaluated similarly:
\begin{eqnarray}
\label{quadrato}
  \Av \left(\Omega (H_\Lambda (\sigma)^2)\right)&=&
  \Av \left(\Omega \left(\sum_{X,Y}J_X J_Y \sigma_X \sigma_Y\right)\right)\nonumber\\
   &=& \Av \left( \sum_{X,Y} J_X J_Y \right)\nonumber\\
   &=& \sum_{X\ne Y} \mu_X \mu_Y +\sum_X (\mu_X^2 +\Delta_X^2)\nonumber\\
   &=& \sum_{X,Y}\mu_X \mu_Y +\sum_X \Delta_X^2
\end{eqnarray}
Therefore, using (\ref{media}) and (\ref{quadrato}) in the
expression for the variance of $h$ one finds
\begin{eqnarray}
\langle h(\sigma)^2 \rangle -\langle h(\sigma )\rangle^2
&=&
\frac{1}{|\Lambda |^2} \Av \left(\Omega (H_\Lambda (\sigma)^2)\right)
   -\frac{1}{|\Lambda |^2} \{\Av \left(\Omega (H_\Lambda (\sigma))\right)\}^2
   \nonumber\\
& = &
\frac{1}{|\Lambda|^2} \sum_X \Delta_X^2 \le \frac{\bar{c}}{|\Lambda|}
\end{eqnarray}
\qed


\vspace{1truecm}
\noindent
{\bf Acknowledgments}. P.C and C.G. thank the Tokyo Institute of Technology for the hospitality during
the period July - August 2007 in which this work was developed and Michel Talagrand for interesting 
discussions.

\end{document}